\documentclass[twocolumn,showpacs,preprintnumbers,amsmath,amssymb]{revtex4}
\usepackage{graphicx}% Include figure files
\usepackage{dcolumn}% Align table columns on decimal point
\usepackage{bm}% bold math

\begin{document}

\title{Observation of a subharmonic gap singularity in interlayer tunneling characteristics of Bi$_2$Sr$_2$CaCu$_2$O$_{8+\delta}$}

\author{V. M. Krasnov}
\email{Vladimir.Krasnov@fysik.su.se}

\affiliation{Department of Physics, Stockholm University, AlbaNova
University Center, SE-10691 Stockholm, Sweden}

\date{\today }

\begin{abstract}

A subharmonic structure in Josephson junctions appears due to
Andreev reflections within the junction. Here we report on
experimental observation of a subharmonic half-gap singularity in
interlayer tunneling characteristics of a layered high temperature
superconductor Bi$_2$Sr$_2$CaCu$_2$O$_{8+\delta}$. The singularity
is most pronounced in optimally doped crystals and vanishes with
decreasing doping. It indicates existence of non-vanishing
electronic density of states and certain metallic properties in
the intermediate BiO layers, which grows stronger with increasing
doping. This provides an additional coherent interlayer transport
channel and can explain a gradual transition from an incoherent
quasi-two-dimensional $c$-axis transport in underdoped to a
coherent metallic transport in overdoped cuprates. Furthermore,
due to a very small sub-gap current, the singularity allows
unambiguous extraction of the superconducting gap, without
distortion by self-heating.


\end{abstract}

\maketitle

The mechanism of interlayer ($c$-axis) transport in cuprate high
temperature superconductors remains an actively debated subject. A
qualitative difference between metallic in-plane and non-metallic
out-of-plane resistivities \cite{Watanabe_1997} is a strong
indication for predominantly incoherent nature of $c$-axis
transport, which is achieved by interlayer hopping or tunneling
\cite{Anderson_1988,McKenzie_1998,Leggett_2001,Carbotte_2001,Ferrero_2010,Hartnoll_2015}.
The tunneling nature of $c$-axis transport leads to appearance of
the intrinsic Josephson effect between CuO$_2$ planes in layered
Bi$_2$Sr$_2$CaCu$_2$O$_{8+\delta}$ (Bi-2212) cuprates at
temperatures below $T_c$ \cite{Kleiner}. However, the electronic
system in cuprates is not strictly two-dimensional. This has been
demonstrated by observation of bonding-antibonding bilayer
splitting of electronic bands \cite{Kordyuk_2002}. Indications for
coherent transport were obtained in strong magnetic fields
\cite{Vignolle_2012}. Since the two dimensional superconductivity
is suppressed by fluctuations \cite{Varlamov_2011}, presence or
absence of the coherent metallic transport in the $c$-axis
direction, i.e., in the third dimension, and the mechanism of
interlayer coupling remain to be important issues for
understanding high temperature superconductivity.

The intrinsic Josephson effect provides an accurate way of probing
weak interlayer coupling in cuprates. Due to a $d$-wave symmetry
of the order parameter, the product of the Josephson critical
current $I_c$ and the normal resistance $R_n$ in intrinsic
junctions should strongly depend on the coherence (momentum
conservation) upon tunneling. The $I_c R_n$ is maximum $\sim
\Delta/e$ for coherent, and zero for completely incoherent
tunneling \cite{Tanaka_1997}. Here $\Delta$ is the maximum value
of the superconducting energy gap. Analysis of $I_c R_n$ in
intrinsic Josephson junctions indicated that in overdoped Bi-2212
interlayer tunneling is predominantly coherent $I_c R_n \sim
\Delta/e$ \cite{Doping,Krasnov_Farc}. However, $eI_c R_n/\Delta$
rapidly decreases upon opening of the pseudogap in the underdoped
state \cite{Doping,Krasnov_Farc}. This may either indicate that
interlayer tunneling becomes progressively more incoherent with
decreasing doping \cite{Doping}, or that the Fermi surface is
reconstructed by the pseudogap \cite{Krasnov_Farc}.

Intrinsic Josephson junctions are characterized by low dissipation
\cite{Katterwe_2010}. This is often taken as evidence for
Superconductor-Insulator-Superconductor (SIS) structure of
Bi-2212, in which S are superconducting CuO$_2$-Ca-CuO$_2$
bilayers and I is the insulating SrO-2BiO-SrO layer. Yet, this
does not preclude that some of the layers in the SrO-2BiO-SrO
stack are metallic, like in case of SINIS (N - is a normal metal)
or SIS'IS junctions
\cite{Brinkman_2000,Cassel_2001,Rajauria_2008}, provided that
transparency of the I interface is sufficiently low. The metallic
behavior of the intermediate layer is manifested in appearance of
subharmonic gap structure
\cite{Octavio_1983,Kleinsasser_1994,Shumeiko_2001} due to Andreev
reflection of quasiparticles into Cooper pairs \cite{Andreev} at
the SIN interface.

In this work we report on observation of the subharmonic gap
singularity in intrinsic tunneling characteristics of small
Bi-2212 mesa structures. This indicates presence of a finite
electronic density of states in BiO layers. The singularity is
most pronounced in optimally doped crystals and decreases with
decreasing doping. The sub-gap singularity allows evaluation of
the gap value at negligible self-heating. We demonstrate that the
energy gap can be confidently extracted for more than two orders
of magnitude variation of the dissipation power. This provides an
important self-consistency check for intrinsic tunneling
spectroscopy.
%and confirms that high-temperature superconductivity above $T_c$ occurs as a result of amplitude .
Finally, we discuss consequences of the metallic behavior of BiO
layers, which provides an additional coherent transport channel
and can explain a gradual transition from a two-dimensional
incoherent to a three-dimensional coherent $c$-axis transport with
increasing doping.

\begin{figure*}
\includegraphics[width=0.9\textwidth]{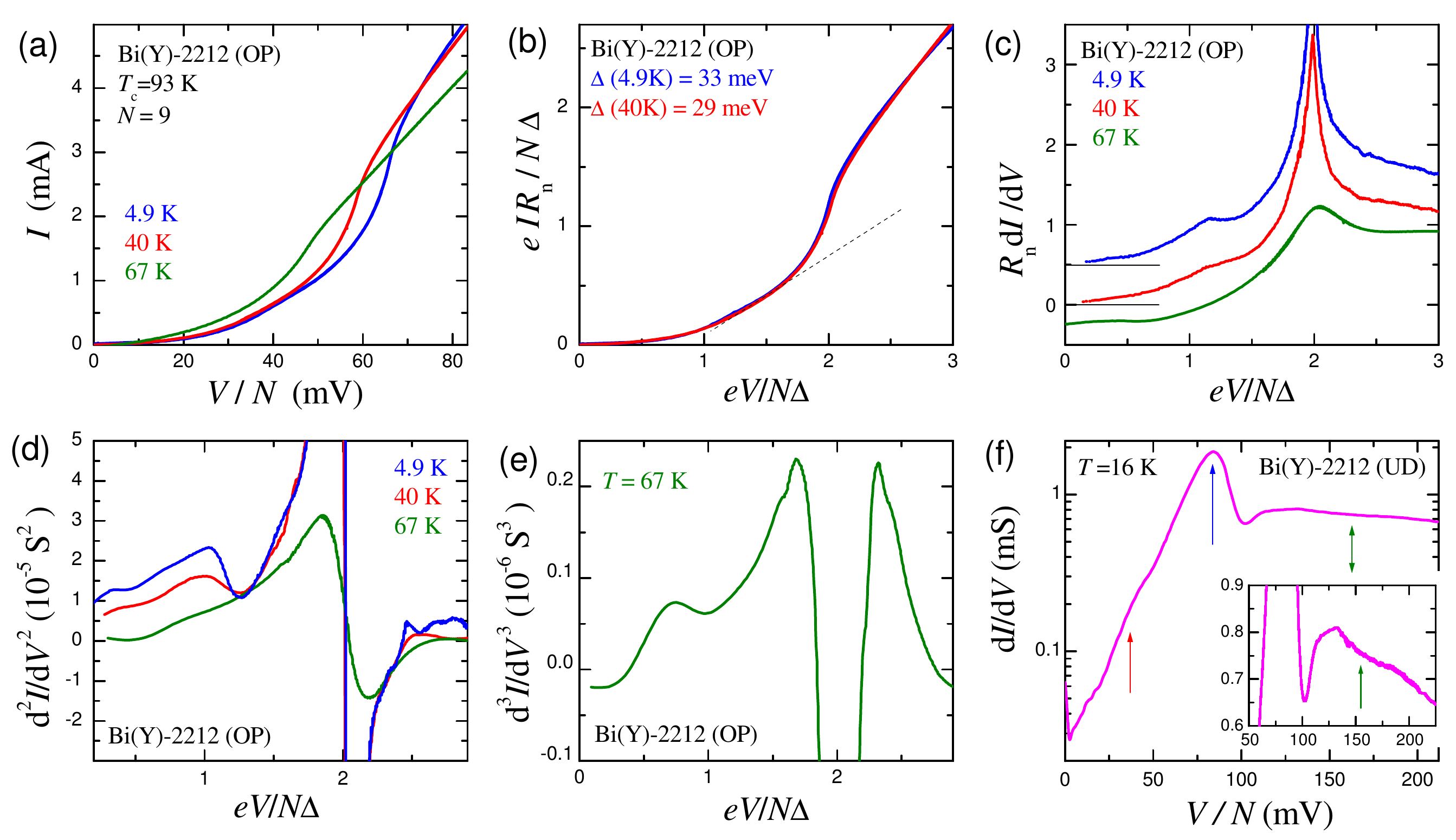}
\caption{(Color online). (a) Current-Voltage characteristics of an
optimally doped (OP) Bi(Y)-2212 mesa at different $T<T_c$. Only
the last quasiparticle branches with all junctions in the
resistive state are shown. (b) The same data, after normalization
by $\Delta(T)$. The collapse of curves into one indicate that both
voltage and current scales are determined by the gap value. Dashed
line indicate a rapid increase of the sub-gap current at
$eV/N>\Delta$. (c) $dI/dV(V)$ and (d) $d^2I/dV^2$ curves for the
same data (the curves are shifted vertically for clarity). (e)
Third-derivative characteristics at $T\simeq 67$ K. (f) The
$dI/dV$ characteristics of a small underdoped (UD) Bi(Y)-2212
mesa. Arrows indicate the subharmonic, sum-gap and four-gap
singularities. }\label{Fig1}
\end{figure*}

We present data for three batches of single crystals: the
Y-substituted Bi$_2$Sr$_2$Ca$_{1-x}$Y$_x$Cu$_2$O$_{8+\delta}$,
Bi(Y)-2212), with the maximum $T_c\simeq 94.5$K; the lead-doped
Bi$_2-x$Pb$_x$Sr$_2$Ca$_1$Cu$_2$O$_{8+\delta}$, Bi(Pb)-2212, with
the maximum $T_c\simeq 93$K, and the pure Bi-2212 with the maximum
$T_c \simeq 86$K. Small mesa structures were made on freshly
cleaved crystals using micro/nano-fabrication techniques. Details
of mesa fabrication and characterization are described elsewhere
\cite{Katterwe_PRL2008,SecondOrder,MR,Jacobs_Bi2201}. All
presented measurements are performed at ambient magnetic field.
The doping state of mesas was determined from a systematic study
of all the characteristics, including (but not only) the $T_c$.
Details, including the raw experimental data for different doping
states can be found in Refs.
\cite{Doping,Krasnov_Farc,Jacobs_2016}. In particular, in Ref.
\cite{Krasnov_Farc} oxygen-doped mesas from the same batch were
studied. It was shown that the critical current of the mesas
strongly (almost exponentially) depends on doping. This provides
an accurate way of determination of doping close to optimal doping
where $T_c$ vs. doping is flat.

Figure \ref{Fig1} (a) shows the current-voltage ($I$-$V$)
characteristics of a near optimally
doped Bi(Y)-2212 mesa %($\sim 4\times 7 \mu m^2$,
with $T_c \simeq 93K$ and $N=9$ intrinsic Josephson junctions, at
$T= 4.9$, 40 and 67 K. A pronounced current step occurs at the
sum-gap voltage $V_{sg}/N = 2\Delta/e$, followed by the ohmic and
almost $T-$independent resistance \cite{SecondOrder}. Such a
behavior is typical for superconducting tunnel junctions
\cite{MR}. Fig. \ref{Fig1} (b) demonstrates that the $I$-$V$
curves at different $T$ merge in one when both the voltage and the
current scales are normalized by $\Delta (T)$. This is expected
for SIS junctions, in which not only the voltage, but also the
current scale is proportional to the superconducting gap, as seen
from theoretical $I$-$V$ curves in Fig. \ref{Fig2} (a). Fig.
\ref{Fig1} (c) shows normalized $dI/dV(V)$ characteristics for the
same mesa. A sharp sum-gap peak occurs at $eV_{sg}/N\Delta =2$.
Simultaneously, we notice an additional bump at a half of the peak
voltage, $eV/N\Delta \simeq 1$. This subharmonic gap feature will
be in focus of this work. The subharmonic feature is rapidly
smeared out with increasing temperature, but can be traced using
higher derivatives $d^2I/dV^2$, Fig. \ref{Fig1} (d), and
$d^3I/dV^3$, Fig. \ref{Fig1} (e).

With decreasing doping the sum-gap peak is decreasing in amplitude
\cite{Doping,Krasnov_Farc}. Fig. \ref{Fig1} (f) shows a $dI/dV$
characteristics of a moderately underdoped Bi(Y)-2212 mesa with
$T_c\simeq 91$ K at $T=16$ K (note the semi-logarithmic scale). It
is seen that for the underdoped mesa both the sum-gap peak and the
subharmonic bump have significantly smaller amplitudes than for
the near optimally doped case, Fig. \ref{Fig1} (c).

\begin{figure}
\includegraphics[width=0.48\textwidth]{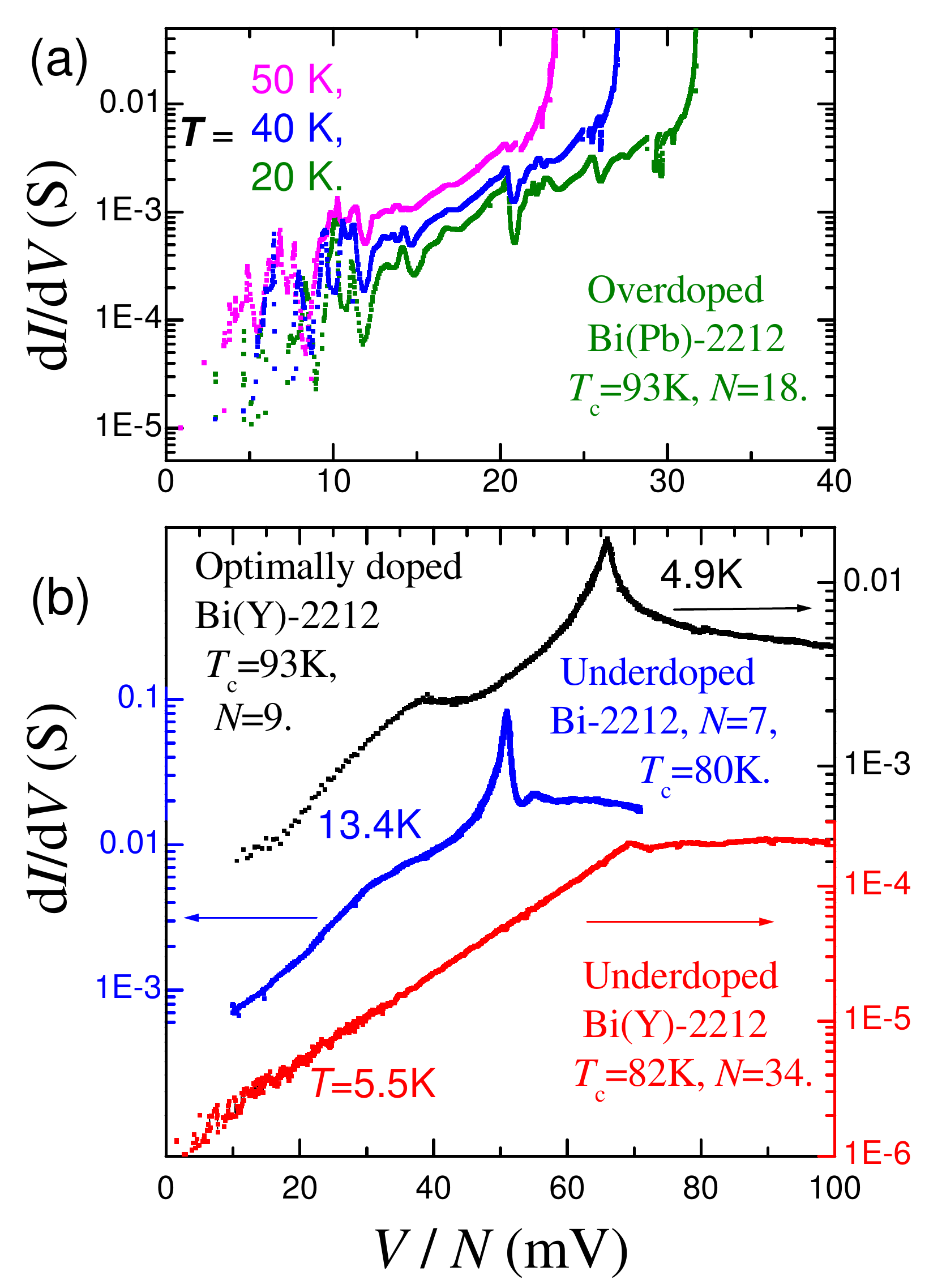}
\caption{(Color online). $dI/dV$ characteristics versus voltage
per junction for mesas with different doping levels. (a) Sub-gap
parts of $dI/dV$ for an overdoped Bi(Pb)-2212 mesa. Strong
singularities are due to phonon resonances. Note that positions of
phonon resonances are temperature independent. (b) $dI/dV$ curves
for near optimally doped Bi(Y)-2212 (black), slightly underdoped
Bi-2212 (blue) and moderately underdoped Bi(Y)-2212 (red) mesas.
Note that the subharmonic singularity disappears with decreasing
doping. }\label{FigS1}
\end{figure}

The sum-gap and the subharmonic singularities are not the only
spectroscopic features in intrinsic tunneling characteristics. A
double-arrow in Fig. \ref{Fig1} (f) points at a small dip in
conductance, which occurs at approximately twice the sum-gap peak
voltage, i.e., $eV/N \sim 4\Delta$. The inset in Fig. \ref{Fig1}
(f) shows a zoom-in on this feature. As discussed in Refs.
\cite{Cascade,Boson}, this dip is caused by reabsorbtion of
nonequilibrium bosons generated upon relaxation of injected
electrons.

Figure \ref{FigS1} demonstrates doping dependence of the
subharmonic singularity. Fig. \ref{FigS1} (a) shows sub-gap parts
of differential conductance $dI/dV$ vs voltage per junction for an
overdoped Bi(Pb)-2212 mesa. It is seen that in overdoped crystals
the sub-gap conductance is dominated by strong phonon resonances
\cite{Schlenga_1998,Ponomarev,Katterwe_2011}, which appear at
temperature-independent voltages. Presence of phonon resonances
makes it difficult to analyze the subharmonic gap structure. Fig.
\ref{FigS1} (b) shows $dI/dV(V/N)$ curves for optimally doped and
underdoped mesas. It is seen that with decrease of doping the
subharmonic feature is strongly decreased and gets completely
washed away in moderately underdoped crystals
\cite{Katterwe_PRL2008,SecondOrder}, while the sum-gap peak is
still clearly visible. This may indicate changes in interlayer
transport mechanism with doping \cite{Katterwe_PRL2008}.

\begin{figure*}
\includegraphics[width=0.95\textwidth]{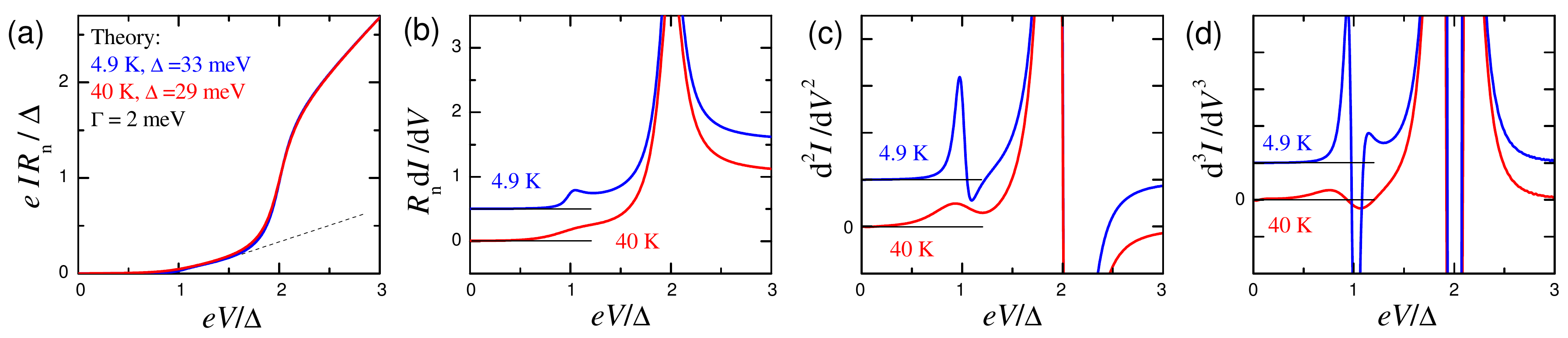}
\caption{\label{Fig2}(Color online). Numerically calculated
characteristics for a gapless SIS junction with a finite depairing
factor $\Gamma=2$ meV and with parameters of optimally doped
Bi(Y)-2212 mesa from Fig. \ref{Fig1}. (a) $I$-$V$ characteristics
at $T=4.9$ and 40 K scaled by $\Delta(T)$. (b), (c) and (d) first,
second and third derivatives (curves for different $T$ are shifted
vertically for clarity). Note appearance of the sub-harmonic
singularity at $eV=\Delta$ due to the gaplessness. }
\end{figure*}

\begin{figure*}[!]
\includegraphics[width=0.9\textwidth]{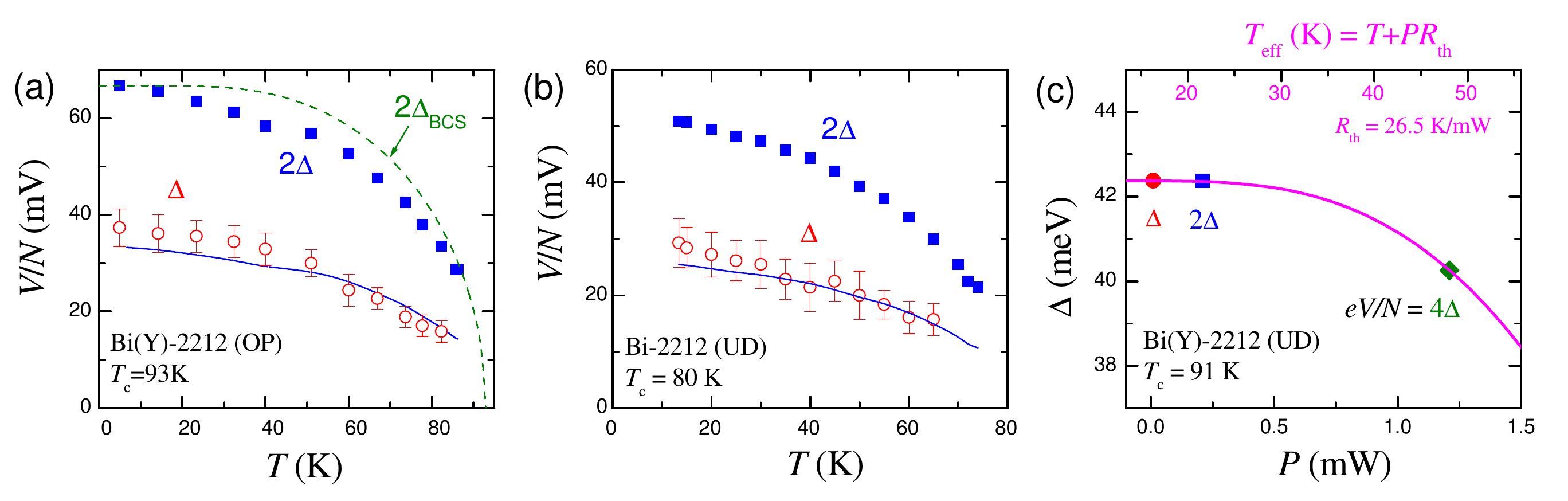}
\caption{\label{Fig3} (Color online). (a) and (b) Temperature
dependencies of the sum-gap (squares) and the sub-gap
singularities for (a) an optimally doped Bi(Y)-2212 mesa from Fig.
\ref{Fig1}, and (b) for an underdoped Bi-2212 mesa. Solid lines
represents half the sum-gap voltage. The dashed line in (a) shows
the conventional BCS $T$-dependence of the superconducting gap.
(c) Dissipation power dependence of the gap extracted from the
sub-harmonic (circle) sum-gap (square) and four-gap (rhombus)
singularities for a small underdoped Bi(Y)-2212 mesa from Fig.
\ref{Fig1} (f). Solid line represents the BCS gap at the effective
mesa temperature.%, raised by self-heating.
}
\end{figure*}

A pronounced subharmonic gap structure at $V_n=2\Delta/e(n+1)$,
$n=1,2,...$ has been observed in superconducting point contacts
\cite{Scheer_2000} and SIS junctions with pinholes
\cite{Kleinsasser_1994}. The subharmonic structure is usually
attributed to multiple Andreev reflections in quasi
one-dimensional quantum channels \cite{Shumeiko_2001}. The
$d$-wave symmetry of the order parameter in cuprates should not
destroy the subharmonic structure, but can affect its shape. In
particular, it can cause an asymmetry between odd and even $n$
singularities as well as certain smearing due to angular
dependence of the gap \cite{Fogelstrom_2002}. The sub-gap
structure has indeed been observed in cuprate junctions
\cite{Deutscher,Koren_2011,Ngai_2007,Iguchi_2003}, although not
all of it could be ascribed to Andreev reflections. For Bi-2212
cuprates the subharmonic gap structure has not been observed sofar
neither in point contacts \cite{Ozyuzer_2000}, nor in intrinsic
junctions, although some excess sub-gap noise was reported for the
latter \cite{Saito_2004}, which might be related to Andreev
reflections \cite{Fogelstrom_2002}.

A single sub-gap feature reported here is hard to reconcile with
one-dimensional pinholes in the barrier, for which one would
expect to see a series of subharmonic singularities
\cite{Kleinsasser_1994}. Rather it closely resembles the
characteristics of homogeneous two-dimensional SINIS junctions
\cite{Brinkman_2000,Cassel_2001,Rajauria_2008}. In SINIS junctions
quasiparticles can travel at arbitrary angles with respect to the
interface. The corresponding angular averaging leads to a
significant smearing of the subharmonic structure so that only a
leading $n=1$ singularity remains distinguishable. Thus,
observation of a single subharmonic feature may indicate presence
of a finite metallic conductivity in BiO layers.

This, however, is not the only possible interpretation. A similar
single subharmonic feature at $eV=\Delta$ occurs also in SIS
junction when S is a gapless superconductor with a finite
electronic density of states at the Fermi level. The gaplessness
can originate from nodes in the gap in combination with partly
incoherent (momentum non-conserving) tunneling and from impurity
scattering. Fig. \ref{Fig2} presents numerical simulations for a
gapless SIS junction, in which the finite density of states at the
Fermi level was introducing adding a depairing factor (inverse
quasiparticle lifetime) $\Gamma=2$ meV to the conventional BCS
density of states (for details see e.g., the Supplementary
material to Ref. \cite{Katterwe_PRL2008}). Fig. \ref{Fig2} (a)
shows calculated $I$-$V$ curves at $T=4.9K$ and $40K$, with both
$I$- and $V$-scales normalized by $\Delta(T)$. Gaps and
temperatures are the same as in Fig. \ref{Fig1} (a) to facilitate
a direct comparison.
%Thus normalized $I-V$ curves merge almost in one because both the position and the amplitude of the sum-gap
%step is proportional to $\Delta$, consistent with Fig. \ref{Fig1} (b).
A single subharmonic singularity can be seen as an approximately
linear upturn of the quasiparticle current at $V \gtrsim
\Delta/e$, indicated by dashed lines in Figs. \ref{Fig1} (b) and
\ref{Fig2} (a). The corresponding $dI/dV(V)$, $d^2I/dV^2$ and
$dI^3/dV^3$ curves, normalized by $\Delta (T)$, are shown in Figs.
\ref{Fig2} (b), (c) and (d), respectively. Those theoretical
curves are consistent with experimental data from Figs. \ref{Fig1}
(c-e). The subharmonic feature in this case is entirely due to
gaplessness and has the same origin as the singularity at
$eV=\Delta$ in SIN junctions.

Thus, a single half-gap singularity is expected both in SINIS
junctions with finite density of states in the intermediate
N-layer and in gapless SIS junctions with finite density of states
at the Fermi level in the S-layers. For our intrinsic junctions
those two cases would correspond to a finite metallic density of
states in BiO layers or to a gapless case with a finite density of
states in CuO$_2$ layers. It is difficult to discriminate those
scenarios just by looking at the shapes of $dI/dV(V)$ curves
because the latter look very similar in both cases. However, a
certain discrimination between those two scenarios can be made
from the analysis of doping evolution of the singularity. As seen
from Fig \ref{FigS1} (b), the subharmonic singularity is rapidly
decreasing with decreasing doping. From angular-resolved
photoemission \cite{Norman_1998,Vishik_2009} and surface tunneling
spectroscopy \cite{Alldredge_2008} it is known that the depairing
factor $\Gamma$ is increasing with decreasing doping. Therefore,
for the gapless SIS junction scenario the residual density of
states at the Fermi level should not decrease with decreasing
doping and the subharmonic singularity should still remain visible
in underdoped junctions, which is not the case. To the contrary,
for the SINIS scenario with BiO being the N-layer, it is expected
that the density of states in the BiO layer will gradually
decrease with decreasing doping and will eventually vanish upon
approaching the insulating state. Consequently, the observed
disappearance of the subharmonic singularity in moderately
underdoped intrinsic junctions is consistent with SINIS case and
implies presence of finite metallic properties in BiO layers.

Figures \ref{Fig3} (a) and (b) show $T-$ dependencies of the
sum-gap voltage (solid squares) and the sub-gap singularity (open
circles) for (a) the optimally doped Bi(Y)-2212 mesa from Fig.
\ref{Fig1} and (b) for a moderately underdoped Bi-2212 mesa with
$T_c\simeq 80 K$, studied in Ref. \cite{Katterwe_PRL2008}. The
solid lines represent half of the sum-gap voltage and demonstrates
that the sub-gap feature indeed represents the subharmonic
half-gap singularity, rather than phonon resonances, which are
$T$-independent \cite{Ponomarev,Katterwe_2011}, as seen from Fig.
\ref{FigS1} (a).
%Remarkably, the singularity is distinguishable in the same $T-$ range as the
%sum-gap peak in $dI/dV$. This indicates that not only the voltage
%but also the amplitudes of the two singularities are correlated.

Observation of the subharmonic singularity allows accurate
extraction of the genuine gap value, not affected by self-heating.
Indeed, due to a smallness of the sub-gap current, the subharmonic
singularity corresponds to a very small dissipation power and,
therefore, is free from self-heating. For example, for the small
Bi(Y)-2212 mesa from Fig. \ref{Fig1} (f), $P=0.01$ mW at the
sub-gap, 0.21 mW at the sum-gap and 1.21 mW at the four gap
singularities. Fig. \ref{Fig3} (c) shows gap values obtained from
those three singularities as a function of the dissipation power.
Due to self-heating \cite{Krasnov_2005,SecondOrder} the gap is
gradually decreasing with increasing $P$ because the effective
temperature of the mesa $T_{eff}=T+R_{th}P$ is elevated above the
base temperature $T$. Here $R_{th}$ is the thermal resistance of
the mesa \cite{Krasnov_2005,SecondOrder}. The solid line in Fig.
\ref{Fig3} (c) represents the BCS $\Delta$ vs $T_{eff}$ dependence
(top axis) obtained using $R_{th}$ as a fitting parameter. The
corresponding $R_{th}=26.5$ K/mW is in good agreement with the
values obtained by in-situ measurements of self-heating
\cite{Krasnov_2005} and from the analysis of size-dependence of
intrinsic tunneling characteristics \cite{SecondOrder} on similar
mesas.

From Fig. \ref{Fig3} (c) it is seen that the energy gap can be
confidently obtained from intrinsic tunneling spectroscopy on
small Bi-2212 mesas within more than two orders of magnitudes of
the dissipation power. In all cases, reported here, self-heating
at the subharmonic singularity is negligible. Therefore, this
singularity provides a decisive self-consistency test and confirms
accurate extraction of the energy gap by intrinsic tunneling
spectroscopy made on small mesas \cite{SecondOrder}.
%From Figs. \ref{Fig3} (a) and (b) it is seen that the superconducting gap
%exhibit a strong $T-$dependence with a clear tendency to vanish at
%$T_c$. The dashed line in Fig. \ref{Fig3} (a) represents BCS
%temperature dependence of the gap. Agreement with experiment
%indicates that at optimal doping the superconductivity in cuprates
%appears as a result of the second-order phase transition, like in
%conventional superconductors \cite{SecondOrder}. In other words,
%the superconductivity at $T>T_c$ is destroyed not only by phase
%fluctuation but by reduction of the amplitude of the order parameter.

To conclude, we reported on observation of subharmonic half-gap
singularity in interlayer tunneling characteristics of Bi-2212
cuprates. The subharmonic singularity allows unambiguous
determination of the energy gap because it occurs at a very small
sub-gap current and negligible self-heating. This is an important
step for development of intrinsic tunneling spectroscopy of
cuprates. We have argued that the subharmonic singularity is a
manifestation of the finite electronic density of states at the
Fermi level in the superconducting state. The observed doping
dependence indicated that the phenomenon is most likely brought
forward by metallic behavior of intermediate BiO layer. The latter
may have a significant influence on properties of layered
cuprates. For example, it is well established that the anisotropy
in cuprates is strongly doping dependent \cite{Watanabe_1997}. For
Bi-2212 it changes from about a million in underdoped to a hundred
in overdoped state. For YBa$_2$Cu$_3$O$_{7-x}$ it changes from
about a hundred in underdoped (which even exhibit the intrinsic
Josephson effect \cite{Nagao_2006}) to about five in a slightly
overdoped case. Such a behavior can be explained by a gradual
enhancement of metallic properties of intermediate layers with
increasing doping, which adds a coherent mechanism to interlayer
transport, and provides a way for establishing a fully coherent
three-dimensional $c$-axis transport in the strongly overdoped
metallic (Fermi liquid) state. To the contrary, in the underdoped
state the metallic behavior of BiO layers becomes weak. When the
sheet resistance exceeds the quantum resistance, $h/e^2$, a
metal-insulator transition takes place due to Coulomb blocking of
transport \cite{Baturina}. The corresponding Coulomb energy can
represent one of the contribution to the $c$-axis pseudogap
phenomenon and is consistent with recent observation of an
additional "dressed" electron energy in interlayer tunneling
characteristics of underdoped Bi-2212 \cite{Jacobs_2016}.

\acknowledgements The work was supported by the Swedish Foundation
for International Cooperation in Research and Higher Education
Grant No. IG2013-5453 and the Swedish Research Council Grant No.
621-2014-4314. I am are grateful to S.O. Katterwe for assistance
in experiment.
%and to A. Yurgens, and T. Benseman for providing Bi(Y)-2212 and
%Bi-2212 single crystals, respectively.

%\end{multicols}

\begin{thebibliography}{99}
%\begin{references}

\bibitem{Watanabe_1997} T. Watanabe, T. Fujii, and A. Matsuda, {\em Phys. Rev. Lett.} {\bf 79}, 2113 (1997).
%Anisotropic Resistivities of Precisely Oxygen Controlled Single-Crystal Bi2Sr2CaCu2O81d: Systematic Study on ''Spin Gap'' Effect

\bibitem{Anderson_1988} P.W. Anderson and Z.Zou, {\em Phys. Rev. Lett. } {\bf
60}, 132 (1988).
%"Normal" Tunneling and "Normal" Transport: Diagnostics for the Resonating-Valence-Bond State.

\bibitem{McKenzie_1998} R.H. McKenzie and P.Moses, {\em Phys. Rev. Lett.} {\bf 81}, 4492 (1998).
%Incoherent Interlayer Transport and Angular-Dependent Magnetoresistance Oscillations in Layered Metals

\bibitem{Leggett_2001} M. Turlakov and A.J. Leggett, {\em Phys. Rev. B} {\bf 63}, 064518
(2001).
%Interlayer c-axis transport in the normal state of cuprates

\bibitem{Carbotte_2001} W. Kim and J.P. Carbotte, {\em Phys. Rev. B} {\bf 63}, 054526
(2001).
%Interlayer coupling and c-axis quasiparticle transport in high-Tc cuprates

\bibitem{Ferrero_2010} M. Ferrero, O. Parcollet, A. Georges, G. Kotliar, and D. N. Basov, {\em Phys. Rev. B} {\bf 82},
054502 (2010).
%Interplane charge dynamics in a valence-bond dynamical mean-field theory of cuprate superconductors

\bibitem{Hartnoll_2015} S.A. Hartnoll, {\em Nature Phys.} {\bf 11}, 54 (2015)
%; DOI: 10.1038/NPHYS3174

\bibitem{Kleiner} R. Kleiner and P. M\"{u}ller, {\em Phys. Rev. B} {\bf 49}, 1327
(1994).

\bibitem{Kordyuk_2002} A. A. Kordyuk, S. V. Borisenko, T. K. Kim, K. A. Nenkov, M. Knupfer, J. Fink, M.
S. Golden, H. Berger, and R.Follath,
%Origin of the Peak-Dip-Hump Line Shape in the Superconducting-State (pi;0) Photoemission Spectra of Bi2Sr2CaCu2O8
{\em Phys. Rev. Lett.} {\bf 89}, 077003 (2002).

\bibitem{Vignolle_2012} B. Vignolle, B.J. Ramshaw, J. Day, D. LeBoeuf, S. Lepault, R. Liang, W.N. Hardy, D.A. Bonn, L. Taillefer, and C. Proust,
{\em Phys. Rev. B} {\bf 85}, 224524 (2012).
%Coherent c-axis transport in the underdoped cuprate superconductor YBa2Cu3Oy

\bibitem{Varlamov_2011} A. Glatz, A. A. Varlamov, and V. M. Vinokur, {\em Phys. Rev. B} {\bf 84},
104510 (2011).

\bibitem{Tanaka_1997} Y. Tanaka and S. Kashiwaya, {\em Phys. Rev. B} {\bf 56}, 892
(1997).

\bibitem{Doping} V.M.Krasnov,
{\em Phys. Rev. B} {\bf 65}, 140504(R) (2002).

\bibitem{Krasnov_Farc} V.M. Krasnov,
%Superconducting condensate residing on small Fermi pockets in underdoped cuprates
{\em Phys. Rev. B} {\textbf 91}, 224508 (2015).

\bibitem{Katterwe_2010} S. O. Katterwe, A.Rydh, H. Motzkau, A.B. Kulakov, and V. M. Krasnov {\em Phys. Rev. B} {\bf 82}, 024517 (2010).
%Superluminal geometrical resonances observed in Bi2Sr2CaCu2O8+x intrinsic Josephson junctions

\bibitem{Brinkman_2000} A.Brinkman, and A.A. Golubov, {\em Phys. Rev. B} {\textbf 61}, 11297 (2000).
%Coherence effects in double-barrier Josephson junctions

\bibitem{Cassel_2001} D. Cassel, G. Pickartz, M. Siegel, E. Goldobin, H.H. Kohlstedt, A. Brinkman,
A.A. Golubov, M. Yu. Kupriyanov, and H. Rogalla, {\em Physica C}
{\bf 350}, 276-290 (2001).

\bibitem{Rajauria_2008} S. Rajauria, P. Gandit, T. Fournier, F.W. J. Hekking, B. Pannetier, and H. Courtois, {\em Phys. Rev. Lett.} {\bf 100}, 207002 (2008).
%Andreev Current-Induced Dissipation in a Hybrid Superconducting Tunnel Junction

\bibitem{Octavio_1983} M. Octavio, M. Tinkham, G. E. Blonder, and T. M. Klapwijk,
{\em Phys. Rev. B} {\bf 27}, 6739 (1983)

\bibitem{Kleinsasser_1994} A.W. Kleinsasser, R.E. Miller, W.H. Mallison, and G.B. Arnold,
{\em Phys. Rev. Lett.} {\bf 72}, 1738 (1994).

\bibitem{Shumeiko_2001} T. Lofwander, V.S. Shumeiko, and G. Wendin,
{\em Supercond. Sci. Technol.} {\bf 14}, R53–R77 (2001).
%Andreev bound states in high- T c superconducting junctions

\bibitem{Andreev}A. F. Andreev, {\em Sov. Phys. JETP} {\bf 19}, 1228 (1964)
%THE THERMAL CONDUCTIVITY OF THE INTERMEDIATE STATE IN SUPERCONDUCTORS
% ——1965 Zh. Eksp. Teor. Fiz. 49 655 (Engl. Transl. Sov. Phys.–JETP 22 455 (1966))
%ELECTRON SPECTRUM OF INTERMEDIATE STATE OF SUPERCONDUCTORS

\bibitem{Katterwe_PRL2008} S.O. Katterwe, A.Rydh, and V. M. Krasnov, {\em Phys. Rev. Lett.} {\bf 101}, 087003 (2008).

\bibitem{SecondOrder} V. M. Krasnov,
%Temperature dependence of the bulk energy gap in underdoped Bi2Sr2CaCu2O8+d: Evidence for the mean-field superconducting transition,
{\em Phys. Rev. B} {\bf 79}, 214510 (2009).

\bibitem{MR} V. M. Krasnov, H. Motzkau, T. Golod, A. Rydh, S. O.
Katterwe, and A. B. Kulakov, {\em Phys. Rev. B} {\bf 84}, 054516
(2011).
%Comparative analysis of tunneling magnetoresistance in
%low-Tc Nb/Al-AlOx/Nb and high-Tc Bi2.yPbySr2CaCu2O8+ä intrinsic Josephson junctions

\bibitem{Jacobs_Bi2201} Th. Jacobs et.al.,
%, S. O. Katterwe, H. Motzkau, A. Rydh, A. Maljuk, T. Helm, C. Putzke, E. Kampert, M. V. Kartsovnik, and V. M. Krasnov
%Electron-tunneling measurements of low-Tc single-layer Bi2+xSr2.yCuO6+�: Evidence for a scaling disparity between superconducting and pseudogap states,
{\em Phys. Rev. B} {\bf 86}, 214506 (2012).

%\bibitem{Suppl} See EPAPS Document No.XXX.
%The supplementary figure demonstrates doping dependence of the subharmonic feature.

\bibitem{Schlenga_1998} K.Schlenga et al., {\em Phys. Rev. B} {\bf 57}, 14518 (1998)

\bibitem{Ponomarev} Ya.G. Ponomarev et al.,
%E.B. Tsokur, M.V. Sudakova, S.N.Tchesnokov, M. E. Shabalin, M. A. Lorenz, M. A. Hein, G. M¨uller, H. Piel, and B. A. Aminov,
{\em Solid State Commun.} {\bf 111}, 513 (1999).

\bibitem{Katterwe_2011} S.O. Katterwe, H. Motzkau, A. Rydh, and
V.M. Krasnov, {\em Phys. Rev. B} {\bf 83}, 100510(R) (2011).

\bibitem{Cascade} V.M. Krasnov, {\em Phys. Rev. Lett.} {\bf 97}, 257003
(2006).

\bibitem{Boson} V. M. Krasnov, S.O. Katterwe1, and A. Rydh, {\em Nature
Commun.} {\bf 4}, 2970 (2013); DOI: 10.1038/ncomms3970.

\bibitem{Scheer_2000} E. Scheer, J.C. Cuevas, A. Levy Yeyati, A. Martin-Rodero, P. Joyez, M.H.
Devoret, D. Esteve, C. Urbina, {\em Physica B} {\bf 280}, 425
(2000).
%Conduction channels of superconducting quantum point contacts.

\bibitem{Fogelstrom_2002} A. Poenicke, J. C. Cuevas, and M.
Fogelstr\"{o}m, {\em Phys. Rev. B} {\bf 65}, 220510(R) (2002).
%Subharmonic gap structure in d-wave superconductors.

\bibitem{Deutscher} G. Deutscher, {\em Nature} {\bf 397}, 410 (1999).

\bibitem{Koren_2011} G. Koren and T. Kirzhner,
{\em Phys. Rev. B} {\bf 84}, 134517 (2011).
%Transport and spectroscopic properties of superconductor/ferromagnet/superconductor junctions of La1.9Sr0.1CuO4/La0.67Ca0.33MnO3/La1.9Sr0.1CuO4

\bibitem{Ngai_2007} J.H. Ngai, W.A. Atkinson, and J.Y.T. Wei,
{\em Phys. Rev. Lett.} {\bf 98}, 177003 (2007).
%Tunneling Spectroscopy of c-Axis Y1-xCaxBa2Cu3O7-d Thin-Film Superconductors

\bibitem{Iguchi_2003} T. Imaizumi, T. Kawai, T. Uchiyama, and I. Iguchi,
{\em J. Low Temp. Phys.} {\bf 131}, 809 (2003).
%Tunnel junctions using different high-Tc superconductors.

\bibitem{Ozyuzer_2000} L. Ozyuzer, J.F. Zasadzinski, C. Kendziora, and K.E. Gray, {\em Phys. Rev. B} {\bf 61}, 3629 (2000).
%Quasiparticle and Josephson tunneling of overdoped Bi2Sr2CaCu2O8+d single crystals.

\bibitem{Saito_2004} A. Saito, H. Ishida and K. Hamasaki, A. Irie and G.
Oya, {\em Appl. Phys. Lett.} {\bf 85}, 1196 (2004).
%Measurements of low-frequency noise spectral densities for a small-sized stack of intrinsic Josephson junctions of Bi 2 Sr 2 CaCu 2 O y single crystal

\bibitem{Norman_1998} M. Norman, M. Randeira, H. Ding, and J. C. Campuzano,
{\em Phys. Rev. B} {\bf 57}, R11093 (1998).
%Phenomenology of the low-energy spectral function in high-Tc superconductors

\bibitem{Vishik_2009} I. M. Vishik, E. A. Nowadnick, W. S. Lee, Z. X. Shen, B. Moritz, T. P. Devereaux, K. Tanaka, T. Sasagawa and T. Fujii,
{\em Nature Phys.} {\bf 5}, 718 (2009).
%A momentum-dependent perspective on quasiparticle interference in Bi2Sr2CaCu2O8+d

\bibitem{Alldredge_2008} J.W. Alldredge et al.,
J. Lee, K. McElroy, M. Wang, K. Fujita, Y. Kohsaka, C. Taylor, H.
Eisaki, S. Uchida, P.J. Hirschfeld and J.C. Davis, {\em Nature
Phys.} {\bf 4}, 319 (2008).
%Evolution of the electronic excitation spectrum with strongly diminishing hole density in superconducting Bi2Sr2CaCu2O8+d

\bibitem{Krasnov_2005} V.M. Krasnov, M. Sandberg, and I. Zogaj, {\em Phys. Rev. Lett.} {\bf 94}, 077003 (2005). %Physica C {\bf 372-376}, 103 (2002)

%\bibitem{Halbritter_1998} J. Halbritter, {\em Physica} (Amsterdam) {\bf 302C}, 221 (1998).

\bibitem{Baturina} B. Sacepe, C. Chapelier, T.I. Baturina, V.M. Vinokur, M.R. Baklanov, and M. Sanquer,
{\em Nature Commun.} {\bf 1}, 140 (2010).

\bibitem{Nagao_2006} M. Nagao, S. Urayama, S. M. Kim, H. B. Wang, K. S. Yun, Y. Takano, T. Hatano, I. Iguchi, T. Yamashita,
M. Tachiki, and H. Maeda, {\em Phys. Rev. B.} {\bf 74}, 054502
(2006).
%Periodic oscillations of Josephson-vortex flow resistance in oxygen-deficient YBa2Cu3Ox.

\bibitem{Jacobs_2016} Th. Jacobs, Y. Simsek, Y. Koval, P. M\"{u}ller, and
V. M. Krasnov, {\em Phys. Rev. Lett.} {\bf 116}, 067001 (2016).
% Sequence of quantum phase transitions in Bi2Sr2CaCu2O8+d cuprates revealed by in-situ electrical doping of one and the same sample



%\bibitem{Golubov_1995} A. A. Golubov et al., {\em Phys. Rev. B} {\bf 51}, 1073 (1995).
%\bibitem{Giaever} I. Giaever and K. Megerle, {\em Phys. Rev.} {\bf 122}, 1101-1111 (1961).
%\bibitem{ARPESreview} A. Damascelli, Z.Hussain, and Z.X.Shen, {\em Rev. Mod. Phys} {\bf 75}, 473 (2003).
%\bibitem{Hamidian_2015} M. H. Hamidian et.al.,
%S. D. Edkins, C. K. Kim, J. C. Davis, A. P. Mackenzie, H. Eisaki, S. Uchida, M. J. Lawler, E.-A. Kim, S. Sachdev and K. Fujita,
%{\em Nature Phys.} {\bf ???} ??? (2015), DOI: 10.1038/NPHYS3519.
%\bibitem{Fujita_2014} K. Fujita et al.,
%M. H. Hamidian, S. D. Edkins, C. K. Kim, Y. Kohsaka, M. Azuma, M. Takano, H. Takagi, H. Eisaki,
%S. Uchida, A. Allais, M. J. Lawler, E.-A. Kim, S. Sachdev, and J. C. Seamus Davis
%{\em PNAS} {\bf 111}, E3026 (2014).
%\bibitem{Wolf_1992} A. Chang, Z.Y. Rong, Yu.M. Ivanchenko, F. Lu, and E. L. Wolf, {\em Phys. Rev. B} {\bf 46}, 5692 (1992).
%Observation of large tunneling-conductance variations in direct mapping of the energy gap of single-crystal Bi2Sr2CaCu2O8-x
%\bibitem{Cren_2000} T. Cren, D. Roditchev, W. Sacks, J. Klein, J.-B. Moussy, C. Deville-Cavellin, and M. Lagu\"{e}s {\em Phys. Rev. Lett.} {\bf 84}, 147 (2000).
%Influence of Disorder on the Local Density of States in High-Tc Superconducting Thin Films
%\bibitem{Misra_2002} S. Misra, S. Oh, D. J. Hornbaker, T. DiLuccio, J. N. Eckstein,
%and A. Yazdani, {\em Phys. Rev. Lett.} {\bf 89}, 087002 (2002).
%Atomic Scale Imaging and Spectroscopy of a CuO2 Plane at the Surface of Bi2Sr2CaCu2O8+d
%\bibitem{KrTemp} V.M.Krasnov, et al.,
%A.Yurgens, D.Winkler, P.Delsing and T.Claeson,
%Phys. Rev. Lett. {\bf 84}, 5860 (2000)
%\bibitem{KrMag} V.M.Krasnov, A.E.Kovalev, A.Yurgens, and D.Winkler, Phys.Rev.Lett. {\bf 86}, 2657 (2001).
%\bibitem{Suzuki_2000}
%M. Itoh, S. I. Karimoto, K. Namekawa, and M. Suzuki, {\em Phys. Rev. B} {\bf 55}, R12001 (1997);
%M. Suzuki and T. Watanabe, {\em Phys. Rev. Lett.} {\bf 85}, 4787 (2000).
%\bibitem{Yurgens_2003} A. Yurgens, D. Winkler, T. Claeson, S. Ono,
%and Y. Ando, {\em Phys. Rev. Lett.} {\bf 90}, 147005 (2003).
%A.V. Puchkov, D. N. Basov, and T. Timusk, J. Phys. Cond. Mat. {\bf 8}, 10049 (1996)
%\bibitem{BCSE} for review see e.g., J.P.Carbotte, Rev. Mod. Phys. {\bf 62}, 1027 (1990)
%\bibitem{TallonPhC} J.L.Tallon and J.W.Loram, Physica C {\bf 349}, 53 (2001)
%\bibitem{Ong} Y.Wang et al., Phys. Rev. Lett. {\bf 95}, 247002 (2005)
%\bibitem{Varlamov} A.Larkin and A.Varlamov
%\bibitem{Fluct1} M.J.Naughton, Phys. Rev. B {\bf 61}, 1605 (2000);
%\bibitem{Fluct2} B.Rosenstein et al, Phys. Rev. B {\bf 63}, 134501 (2001);
%F.P.J.Lin and B.Rosenstein, Phys. Rev. B {\bf 71}, 172504 (2005)
%\bibitem{Landau} I.L.Landau and H.R.Ott, J. Low Temp. Phys. {\bf 139}, 175 (2005); Phys. Rev. B {\bf 66}, 144506 (2002).
%\bibitem{Renner} Ch.Renner et al., Phys.Rev.Lett. {\bf 80}, 149 (1998).
%\bibitem{Yazdani} A.N.Pasupathy et al., Science {\bf 320}, 196 (2008).
%\bibitem{Raman} M. LeTacon, A.Sacuto, A.Georges, G.Kotliar, Y.Gallais, D.Colson, A.Forget,
%Nature Phys. 2, 537 (2006)
%\bibitem{Lee2007} W.S. Lee, et al., Nature {\bf 450}, 81 (2007);
%T.Kondo, T.Takeuchi, A.Kaminski, S.Tsuda, and S.Shin, Phys. Rev. Lett. 98, 267004 (2007).
%\bibitem{Hc2} Y.Wang and H.H.Wen, Euro Phys. Lett. {\bf 81}, 57007 (2008);
%H.Gao et al., Phys. Rev. B {\bf 74}, 020505(R) (2006)
%\bibitem{Ding} H.Ding et al., Phys. Rev. Lett. {\bf 87}, 227001 (2001).
%\bibitem{ARPESp} K.Tanaka et al., Science {\bf 314}, 1910 (2006);
%T.Valla et al., ibid. {\bf 314}, 1914 (2006).
%\bibitem{VanAlphen} N.Doiron-Leyraud et al., Nature {\bf 447}, 565 (2007);
%E.A.Yelland et al., Phys. Rev. Lett. {\bf 100}, 047003 (2008);
%\bibitem{Precursor} V.J.Emery and S.A.Kivelson, Nature {\bf 374}, 434 (1995);
%A.K.Nguyen and A.Sudbo, Phys. Rev. B {\bf 60}, 15307 (1999)
%\bibitem{AF} V.J.Emery, S.A.Kivelson and O.Zachar, Phys. Rev.B {\bf 56}, 6120 (1997).
%\bibitem{Suzuki}
%M.Suzuki, S.Karimoto, and K.Namekawa, J.Phys.Soc.Jap. {\bf 67}, 732 (1998);
%Y.Yamada, et al., Phys.Rev.B {\bf 68}, 054533 (2003)
%\bibitem{Lee} M.H.Bae, et al.,
%J.H.Park, J.H.Choi, H.J.Lee, and K.S.Park,
%Phys. Rev. B {\bf 77}, 094519 (2008).
%\bibitem{Latyshev} Yu.I. Latyshev, et al., Phys. Rev. Lett. {\bf 96}, 116402 (2006)
%\bibitem{Rowell} J.M. Rowell and W.L. Feldmann, Phys. Rev. {\bf 172}, 393 (1968).
%\bibitem{SubHarm} I. Giaever and H.R. Zeller, Phys.Rev.B {\bf 1}, 4278 (1970)
%\bibitem{QCP} C.M.Varma, Phys.Rev.B {\bf 73}, 155113 (2006);
%\bibitem{HeatCom} V.M.Krasnov, Phys.Rev.B {\bf 75}, 146501 (2007)
%\bibitem{KrPhysC} V.M.Krasnov, Physica C {\bf 372-376}, 103 (2002)
%\bibitem{Yamada} Y.Yamada and M.Suzuki, Phys.Rev. B {\bf 66}, 132507 (2002)
%\end{references}
\end{thebibliography}
\end{document}